\documentclass[12pt]{article}
\usepackage{amsmath}
\usepackage{amssymb}
\usepackage{bm}
\usepackage{bbold}
\usepackage{graphicx}
\usepackage{enumerate}
\usepackage{natbib}
\usepackage{hyperref}
\usepackage{booktabs}
\usepackage{url} 
\usepackage{caption}
\usepackage{tikz}
\usetikzlibrary{calc,topaths}
\usepackage{physics}
\usepackage[fontsize=10.5pt]{fontsize}

\begin{document}

\title{Hyperevent network modelling of partially observed gossip data}
\author{Veronica Poda$^a \footnote{veronica.poda@unitn.it}$, Veronica Vinciotti$^a \footnote{veronica.vinciotti@unitn.it}$, Ernst C. Wit$^b \footnote{ernst.jan.camiel.wit@usi.ch}$ \\ \\
\small{$a.$ Department of Mathematics, University of Trento,} \\
\small{$b.$ Institute of Computing, Universit\`a della Svizzera italiana}
}
\date{\today}
\maketitle

\begin{abstract}{
Gossiping is a widespread social phenomenon that shapes relationships and information flow in communities. From a network theoretic point of view, gossiping can be seen as a higher-order interaction, as it involves at least two persons talking about a non-present third. The mechanism of gossiping is complex: it is most likely dynamic, as its intensity changes over time, and possibly viral, if a gossiping event induces future gossiping, such as a repetition or retaliation. We define covariates of interest for these effects and propose a relational hyperevent model to study and quantify these complex dynamics. We consider survey data collected yearly from 44 secondary schools in Hungary. No information is available about the exact timing of the events nor about the aggregate number of events within the yearly time interval. What is measured is whether at least one gossiping event has occurred in a given time interval. We extend inference for relational hyperevent models to the case of right-censored interval-time data and show how flexible and efficient generalized additive models can be used for estimation of effects of interest. Our analysis on the school data illustrates how a model that accounts for linear, smooth and random effects can identify the social drivers of gossiping, while revealing complex temporal dynamics.}
\end{abstract}


\section{Introduction}

Gossiping is a central aspect of social interactions, influencing relationships and affecting group behavior \citep{behaviorgossip,foster_gossip}.  It plays a key role in the regulation of reputations, the reinforcement of group norms, and the dissemination of information within social networks. While it can foster cooperation, it can also lead to harmful outcomes, such as social exclusion and reputational damage \citep{KisfalusiGossipReputation}. Moreover, in a school setting, gossiping can manifest itself as a form of bullying, shaping students’ social and academic development \citep{bullying, bullies}. Understanding the mechanisms that generate and sustain gossiping is therefore relevant to sociology and educational research.

A number of papers have studied the factors that can explain the occurrence of gossiping. These can be about characteristics of the people involved in gossiping. For example,  \citet{nynkegossip} find that gossipers tend to gossip more about a target of their same gender. Other, possibly time-varying, exogenous variables may also have an association with gossiping.  However, the occurrence of gossiping can depend also on the history of past gossiping, which is the case of endogenous factors.   For example, \citet{KisfalusiGossipReputation} find a positive effect for outdegree and indegree popularity. For the first one, this suggests that the occurrence of gossiping tends to induce further occurrences of gossiping in the future, while, for the second one, this means that people who are the target of gossiping may, as a result of this, be the subject of further gossiping. Similarly, reciprocity -- also referred to as retaliation in this setting -- is found by both studies to be present, with the gossiper becoming itself the future target of a gossiping initiated by its original target. Finally, self-reported gossiping seems to co-evolve with perceived gossiping \citep{nynkegossip}, or between gossiping and  competition for reputation \citep{KisfalusiGossipReputation}. 

The studies above show how gossiping, as various other social mechanisms, is characterized by a fundamentally viral nature. This is acknowledged by  \citet{KisfalusiGossipReputation} and \citet{nynkegossip} with the inclusion of factors like reciprocity and popularity in  their statistical model. However, both studies fall short in capturing the complexity of the underlying dynamics. In \cite{nynkegossip}, the effects are estimated from data collected at a single time point, while \citet{KisfalusiGossipReputation} considers longitudinal data but includes in the model only linear effects for the covariates. The latter is partly due to the choice of a  Stochastic Actor-Oriented Model (SAOM) \citep{saom}, as complex dynamic models are computationally expensive and difficult to fit within this framework. Alternatively, in a relational event model the effects of potential drivers have been recently extended to random effects \citep{uzaheta2023random,boschialienspecies}, non-linear and time-varying effects \citep{bauer2022smooth,boschialienspecies,lembo25,rutaheterogeneity}, with efficient computational tools available to fit these complex models.

Gossiping is by definition a higher-order interaction, as it involves at least two people talking about a third non-present person \citep{DoresCruz}. The temporal nature of gossiping has motivatived the choice of continuous time dynamic networks models for modelling the gossip process. Since gossiping is a relational event rather than a relational state, relational event models are more appropriate than SAOMs in capturing the underlying dynamics that describe the sequence of the gossiping hyperevents. Relational hyperevent models (RHEM) \citep{rhem} have been recently introduced as an extension of traditional relational event models to the case of higher-order interactions involving multiple senders and/or multiple receivers.  

We consider the longitudinal school survey study by \citet{gossip_dataset}, in which data are collected once every academic year. The data do not contain information about the exact timing of the gossiping events nor about the number of times a group of students has gossiped  about the same target during  the yearly time interval. What can be deduced by the nominations made by a student is that at least one gossiping event of a certain type has occurred in a given time interval. Ignoring this censoring, as in previous studies \citep{nynkegossip,KisfalusiGossipReputation}, can lead to biased estimates of effects. We therefore develop an extension of the inference for relational hyperevent models to the case of right-censored interval-time data. Within this setting, we show how the likelihood can be written as that of a particular type of regression model. Thus, similarly to traditional relational event models, flexible and efficient generalized additive models can be used for the estimation of effects of interest, so that the potentially complex temporal dynamics of gossiping can be recovered from the partially observed data.

The remainder of this paper is organized as follows. The next session introduces the school survey data described in \citet{gossip_dataset} which motivates the methodological development. We then present the RHEM framework for modelling gossiping hyperevents, followed by a section describing the extension of traditional inferential approaches for RHEM to right-censored interval-time data.  We evaluate the performance of the proposed approach via a simulation study and compare different ways of calculating the effect of time-varying covariates in the case of partially observed data. Finally, we present an illustration of the method in the school survey data, and discuss the gossiping dynamics that are inferred from this.

\section{RECENS: a longitudinal school survey study involving gossiping} \label{sec:survey} 

\begin{figure}[t]
    \centering
    \begin{minipage}[b]{0.4\textwidth}
    \centering
    \includegraphics[width=\linewidth]{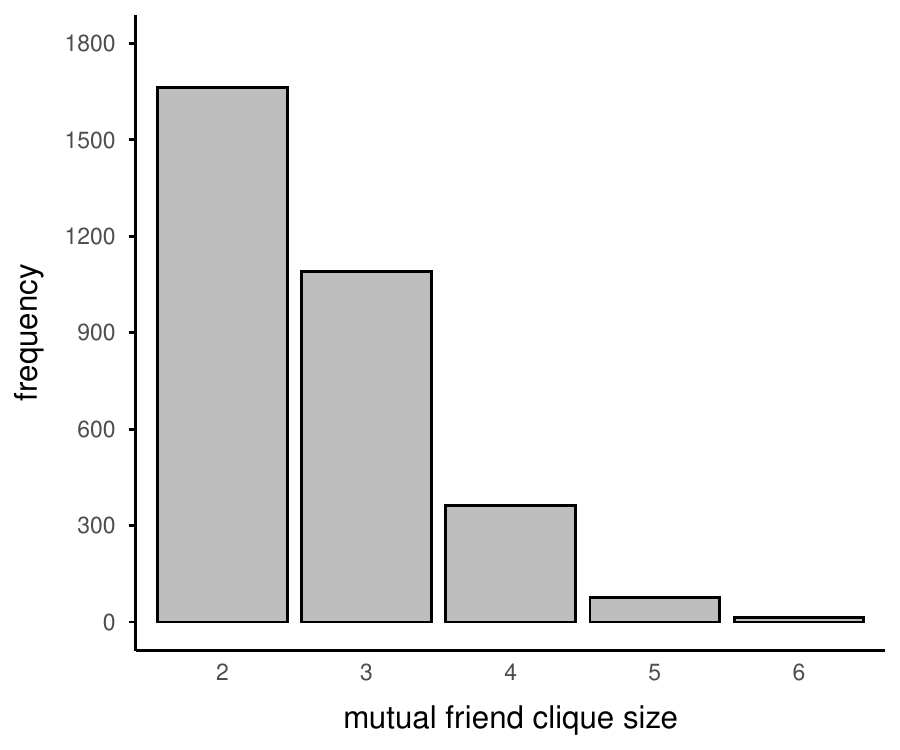} \\
    (a)
    \end{minipage}
    \hfill
    \begin{minipage}[b]{0.4\textwidth}
    \centering
    \includegraphics[width=\linewidth]{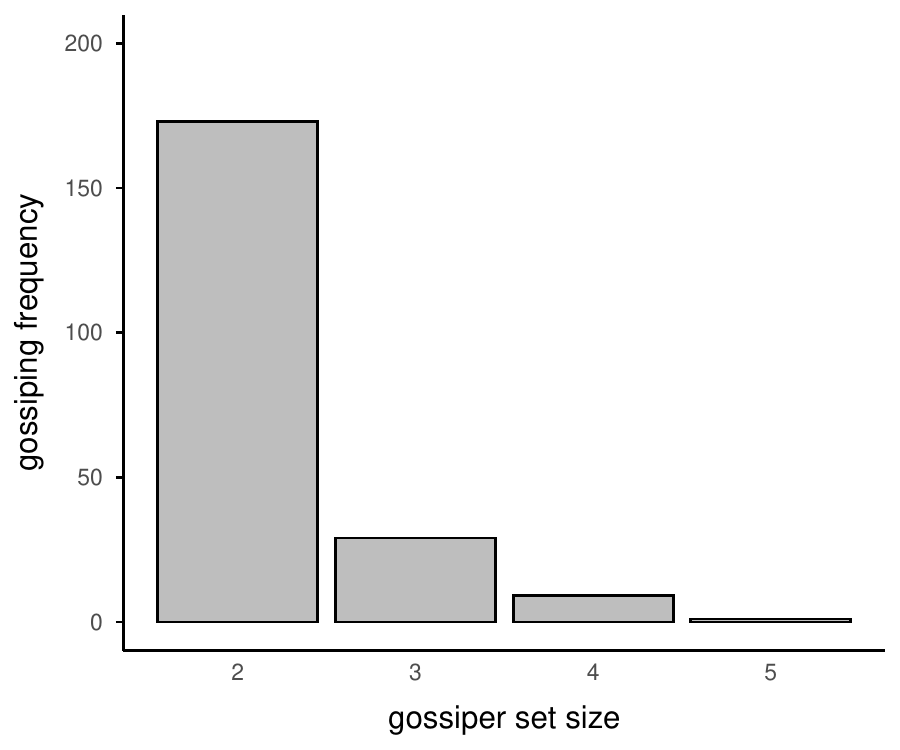} \\
    (b)
    \end{minipage}
        \hfill
        \begin{minipage}[b]{0.4\textwidth}
        \centering
        \includegraphics[width=\linewidth]{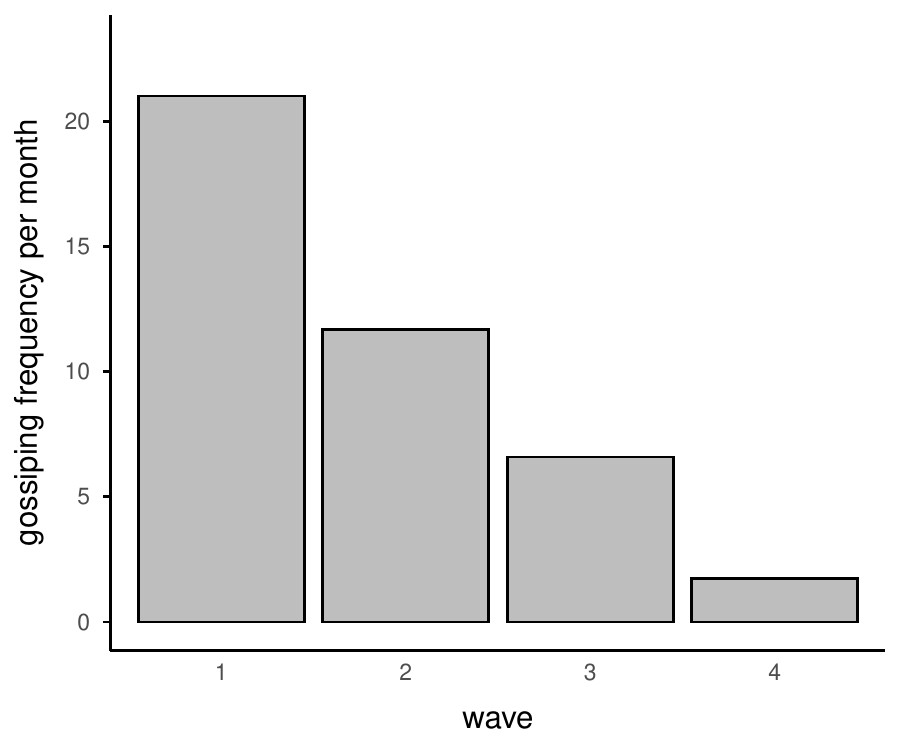}\\
        (c)
    \end{minipage}
    \hfill
    \begin{minipage}[b]{0.4\textwidth}
        \centering            \includegraphics[width=\linewidth]{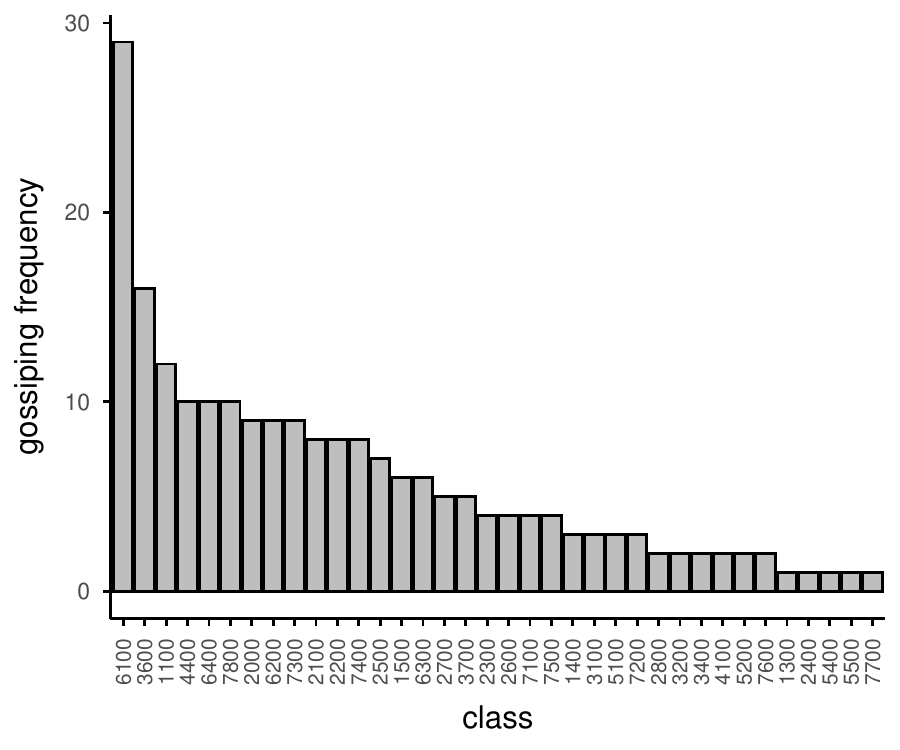} \\
(d)            \end{minipage}

    \caption{(a) Groups of mutual friends tend to be small in size, (b) Gossiping involves predominantly two close friends talking badly about a target,  (c) Gossiping decreases over time,  (d) There is a large heterogeneity in gossiping activity among the classes.}
    \label{fig:gossip_figures}
\end{figure}

In this paper, we study the mechanisms underlying gossiping using  data from the \textit{Wired into Each Other} longitudinal study conducted by the Research Center for Educational and Network Studies (RECENS) in Hungary \citep{gossip_dataset}. The study involves a four-wave survey of $1,686$ students in $44$ Hungarian secondary school classes, across 7 schools in 4 Hungarian towns. The study started with a survey distributed two months after the beginning of high school (October-November 2010) to 9th-grade students enrolled in the selected schools. These are students with an average age of $15.1$ years \citep{gossip_dataset}. Data were further collected six months later (April 2011), one year after the second wave (April 2012), and one year after the third wave (April 2013). The longitudinal nature of the data provides an opportunity to understand the dynamics of gossiping, but it also comes with several methodological challenges.

The first challenge is that the  interaction between gossipers talking badly about someone not present is not observed directly. Instead, the survey records the nominations of each student in response to the question: \textit{Of whom do you say bad things to your friends?} This type of data provides dyadic information that can be used as a proxy to understand gossiping, as it was done in previous studies \citep{KisfalusiGossipReputation,nynkegossip}. In this paper, however, we combine this information with another survey question, where students are asked to nominate their friends: \textit{Please tell us how much you like or dislike your classmates. [–2: strong dislike or hate, –1: dislike, 0: neutral, 1: like, 2: close friendship.]} We use this information, to construct all cliques of mutual close friends within each class. 
Figure~\ref{fig:gossip_figures}a shows how the largest group of mutual friends is made up of $6$ individuals, while the majority of groups are smaller in size.  From this, we define a gossiping hyperevent as the largest group of reciprocated close friendships who talk badly about the same person. In total, we find $212$ of such gossiping hyperevents. Figure~\ref{fig:gossip_figures}b shows how the majority of gossiping hyperevents involve two friends talking badly about a target.

The gossiping hyperevents can be further described by the waves and classes in which they happen, by the receivers and sender groups, and by gender composition. Figure~\ref{fig:gossip_figures}c shows how gossiping decreases over time. At the class level, $9$ out of the $44$ classes recorded no gossiping. Out of the $35$ classes that experienced at least one gossiping hyperevent, Figure~\ref{fig:gossip_figures}d  shows a large heterogeneity in the gossiping activity. 
At the receiver level, $142$ out of the $1,686$ students were targeted by gossiping at least once, 
while $114$ out of the total $7,276$ potential groups of mutual friends participated in at least one gossiping hyperevent.  
This indicates that gossiping is generally rare, which is consistent with previous studies \citep{sparsenet}. In particular, considering all potential sender groups and receivers within each class generates a total of $299,419$ possible gossiping hyperevents across all four waves. Of these,  we observe only $212$. We will later refer to the set of potential events in a wave $k$ as the risk set $\mathcal{R}(k)$.  Finally, as discussed also in the literature \citep{KisfalusiGossipReputation}, the hyperevent data show a predominance of females, both in the role of gossiper and receiver of gossiping. In particular, 80.3\% among gossipers and 74.2\% among receivers  are females. Moreover, females tend to gossip together, as 52.3\% of all sender groups are only female groups, compared to 25.7\% of only males and 21.1\% of mixed groups.

The second challenge that comes with these data is that the nominations made by students only allow us to reconstruct that at least one gossiping event has occurred since the last survey. No information is available about the exact timing of the event nor about the number of times that that gossiping event has occurred. In particular, this means that the $212$ gossiping hyperevents from  Figure~\ref{fig:gossip_figures} are only a lower bound of the total number of gossiping hyperevents that occurred during the four waves. This leads to relational event data that are partially observed both in terms of the event times (interval-censored) and of the counts associated to a gossiping hyperevent (right-censored). Indeed, the information available with a gossiping hyperevent is that there has been at least one occurrence of that event within the wave when it was measured.

In the next sections, we meet these challenges by first describing a relational event model for gossiping hyperevents,  and then by developing an inferential procedure that accounts for the censored nature of the data.

\section{A relational hyperevent model of gossiping}
\label{sec:RHEM}
Since gossiping is a higher-order interaction, we consider a relational hyperevent model \citep{rhem} and define in this section a number of covariates that may be informative in describing the dynamics of gossiping.

\subsection{The relational hyperevent model}

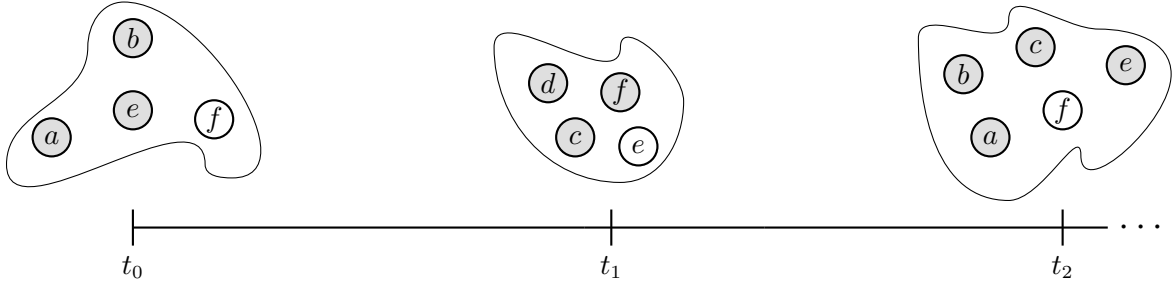
\begin{figure}[t]
    \centering
    \begin{tikzpicture}[scale=1.2]
        \draw[thick] (0,0) -- (5,0) node[right] {};
        \foreach \x/\label in {0/$t_0$, 5.3/$t_1$, 10.3/$t_2$} {
            \draw[thick] (\x,0.2) -- (\x,-0.2) node[below] {\label};
        }

        \draw[thick] (5,0) -- (7,0); 
        \draw[thick] (7,0) -- (10.8,0);
        \node at (11.2,0) {\Large $\cdots$};

\node[above] at (0,3) {\small
  $\begin{array}{rl}
    x_r^{(rd)}(t_0) & = 0\\[5pt]
    x_{Sr}^{(rec)}(t_0) & = 0\\[5pt]
    x_{Sr}^{(sub\_rep)}(t_0) & = 0
  \end{array}$};

\node[above] at (5.3,3) {\small
  $\begin{array}{rl}
    x_r^{(rd)}(t_1) & = 0\\[5pt]
    x_{Sr}^{(rec)}(t_1) & = \frac{1}{3}\\[5pt]
    x_{Sr}^{(sub\_rep)}(t_1) & = 0
  \end{array}$};
  
\node[above] at (10.3,3) {\small
  $\begin{array}{rl}
    x_r^{(rd)}(t_2) & = 1\\[5pt]
    x_{Sr}^{(rec)}(t_2) & = \frac{1}{4}\\[5pt]
    x_{Sr}^{(sub\_rep)}(t_2) & =  \frac{1}{4}
  \end{array}$};

        \node (v1) at (-0.9,1) {$a$};
        \draw[thick] (-0.9,1) circle (6pt); 
        \node (v2) at (0,2.1) {$b$};
        \draw[thick] (0,2.1) circle (6pt); 
        \node (v4) at (0.9,1.2) {$f$};
        \draw[thick] (0.9,1.2) circle (6pt); 
        \node (v5) at (0,1.3) {$e$};
        \draw[thick] (0,01.3) circle (6pt);
        
        \begin{scope}[fill opacity=0.8]
        \filldraw[fill=gray!0] ($(v1)+(-0.5,-0.3)$) 
            to[out=90,in=270] ($(v4) + (-1.4,0.8)$)
            to[out=90,in=180] ($(v2) + (-0.1,0.4)$) 
            to[out=0,in=0] ($(v4) + (0.2,-0.65)$)
            to[out=180,in=270] ($(v2) + (0.8,-1.4)$)
            to[out=90,in=270] ($(v1)+(-0.5,-0.3)$);
        \end{scope}
        
        \foreach \x/\y/\n in {
            -0.9/1/$a$,
            0/2.1/$b$,
            0/1.3/$e$
        }{
            \filldraw[fill=gray!25, thick] (\x,\y) circle (6pt);
            \node at (\x,\y) {\n};
        }
        
        \draw[thick, fill=gray!0] (0.9,1.2) circle (6pt);
        \node (v4) at (0.9,1.2) {$f$};

\begin{scope}[xshift=-1.7cm]

    \node (v1) at (6.3,1.6) {$d$};
    \draw[thick] (6.3,1.6) circle (6pt);
    \node (v2) at (7.1,1.5) {$f$};
    \draw[thick] (7.1,1.5) circle (6pt);
    \node (v3) at (6.6,1) {$c$};
    \draw[thick] (6.6,1) circle (6pt);
    \node (v5) at (7.3,0.9) {$e$};
    \draw[thick] (7.3,0.9) circle (6pt);

    \begin{scope}[fill opacity=0.8]
        \filldraw[fill=gray!0] ($(v1)+(-0.6,0.4)$) 
            to[out=90,in=270] ($(v3) + (0.5,1)$)
            to[out=90,in=90] ($(v5) + (0.5,0.5)$) 
            to[out=270,in=0] ($(v3) + (0.5,-0.5)$)
            to[out=180,in=270] ($(v1) + (-0.6,0.4)$);
    \end{scope}

    \begin{scope}[fill opacity=1] 
        \filldraw[fill=gray!25, thick] (6.3,1.6) circle (6pt);
    \end{scope}
    \node (v1) at (6.3,1.6) {$d$};

    \begin{scope}[fill opacity=1] 
        \filldraw[fill=gray!25, thick] (6.6,1) circle (6pt);
    \end{scope}
    \node (v3) at (6.6,1) {$c$};

    \draw[thick, fill=gray!0] (7.3,0.9) circle (6pt);
    \node (v5) at (7.3,0.9) {$e$};

    \begin{scope}[fill opacity=1] 
        \filldraw[fill=gray!25, thick] (7.1,1.5) circle (6pt);
    \end{scope}
    \node (v2) at (7.1,1.5) {$f$};

\end{scope}

\begin{scope}[xshift=-0.5cm]
    \node (v1) at (10,1) {$a$};
    \draw[thick] (10,1) circle (6pt);
    \node (v2) at (9.7,1.7) {$b$};
    \draw[thick] (9.7,1.7) circle (6pt);
    \node (v3) at (10.5,2) {$c$};
    \draw[thick] (10.5,2) circle (6pt);
    \node (v4) at (11.5,1.8) {$e$};
    \draw[thick] (11.5,1.8) circle (6pt);
    \node (v6) at (10.8,1.3) {$f$};
    \draw[thick] (10.8,1.3) circle (6pt);

    \begin{scope}[fill opacity=0.8]
    \filldraw[fill=gray!0] ($(v2)+(-0.5,0.4)$) 
        to[out=90,in=270] ($(v4) + (-1.3,0.5)$)
        to[out=90,in=180] ($(v3) + (0.7,0.2)$) 
        to[out=0,in=90] ($(v4) + (0.5,-0.1)$)
        to[out=270,in=270] ($(v6) + (0.2,-0.5)$)
        to[out=90,in=0] ($(v1) + (0.2,-0.7)$)
        to[out=180,in=270] ($(v2)+(-0.5,0.4)$);
    \end{scope}

    \begin{scope}[fill opacity=1] 
    \filldraw[fill=gray!25, thick] (10,1) circle (6pt);
    \end{scope}
    \node (v1) at (10,1) {$a$};
    \begin{scope}[fill opacity=1] 
    \filldraw[fill=gray!25, thick]  (9.7,1.7) circle (6pt);
    \end{scope}
    \node (v2) at (9.7,1.7) {$b$};
    \begin{scope}[fill opacity=1] 
    \filldraw[fill=gray!25, thick]  (10.5,2) circle (6pt);
    \end{scope}
    \node (v3) at (10.5,2) {$c$};
    \begin{scope}[fill opacity=1] 
    \filldraw[fill=gray!25, thick]  (11.5,1.8) circle (6pt);
    \end{scope}
    \node (v4) at (11.5,1.8) {$e$};
    \draw[thick, fill=gray!0] (10.8,1.3) circle (6pt);
    \node (v6) at (10.8,1.3) {$f$};
    \end{scope}
    \end{tikzpicture}
    \caption{Higher-order gossiping interactions evolving over time. At each time point, a gossiping hyperevent occurs, with the grey nodes representing the set of gossip senders, while the receiver of the gossip is shown in white. For each hyperevent, we evaluated three covariates from Table \ref{tab:covariates}. At time $t_2$, there is a repetition of the same receiver and a subset repetition with respect to time $t_0$ and a reciprocal event with respect to time $t_1$. }
    \label{fig:hoi}
\end{figure}

A directed relational hyperevent is an interaction between a set of senders and a set of receivers, occurring at a specific point in time. Formally, let \( V \) denote the set of individuals in a social network. A relational hyperevent at time \( t \) is defined as the tuple \( (S, R, t) \), where:
\begin{enumerate}
\item  $S \subset V$ is the set of senders;
\item $R \subset V$ is the set of receivers;
\item $t \in [0,T]$ is the time at which the hyperevent occurs.
\end{enumerate}
In our setting, the receiver set consists of a single individual $r \in V$ who, at time $t$, is the target of a gossip by a sender group $S$ of mutual friends. So we define a gossiping hyperevent as the tuple $(S,r,t)$. A gossiping hyperevent process is a marked point process,
\begin{equation*}
\{((S_k, r_k), \, t_k) \, | \, k \geq 1 \},
\end{equation*}
where, at a random time point $t_k$, an interaction occurs from the set of senders $S_k$ to the receiver $r_k$. Figure~\ref{fig:hoi} provides a visualization of a realization of the stochastic process, with gossiping hyperevents occurring over time.

Associated with the relational event process there exists a multivariate counting process $N$, which records the number of directed interactions from $S$ to $r$ up to time $t$:
\begin{equation*}
N_{Sr}(t) = \sum_{k \geq 1} 1\{t_k \leq t,\ S_k = S,\ r_k = r\}.
\end{equation*}
From the Doob–Meyer decomposition \citep{meyer1962decomposition}, this submartingale process can be split into a predictable process $\Lambda_{Sr}(t)$, and a residual martingale process $M_{Sr}(t)$, that is
\begin{equation*}
N_{Sr}(t)=\Lambda_{Sr}(t)+M_{Sr}(t).
\end{equation*}
If it exists, the derivative of the cumulative hazard process \(\Lambda_{Sr}\)
\begin{equation*}
\lambda_{Sr}(t) = \frac{d \Lambda_{Sr}}{dt} (t)
\end{equation*}
defines the instantaneous hazard for the relational hyperevent \((S, r)\).

A Relational Hyperevent Model (RHEM) describes how covariates of interest are associated to the hazard of a hyperevent. In particular, we consider the model
\begin{equation}
\lambda_{Sr}(t) \, = \, Y_{Sr}(t) \, \lambda_0(t) \,
\exp\!\left\{
f(\boldsymbol{x}_{Sr}(t)) + \boldsymbol{\gamma}^\top \boldsymbol{z}_{Sr}
\right\},\label{hazard}
\end{equation}
with $Y_{Sr}(t)$ an indicator, equal to $1$ if the hyperevent $(S, r)$ is at risk of happening at time $t$, and $0$ otherwise, a non-parametric baseline hazard function $\lambda_0(t)$, which is common to all hyperevents and does not depend on the specific pair $(S, r)$, and a parametric component of the model capturing the effect of covariates of interest on the hazard. The covariates $\boldsymbol{x}_{Sr}(t)$ can include both endogenous and exogenous covariates that are measurable with respect to the history  of the process. In particular, endogenous covariates depend on the history of the relational hyperevent network, such as repetion or reciprocity of a gossiping hyperevent, while exogenous covariates are time-dependent or defined by actor-level attributes but they are independent of previous interactions, such as gender of senders/receiver or their average age.
These covariates can enter the model either linearly, i.e., $f(\boldsymbol{x}_{Sr}(t)) = \boldsymbol{\beta}^\top \boldsymbol{x}_{Sr}(t)$, for some vector of parameters $\boldsymbol{\beta}$, or non-linearly through a more flexible smooth function $f(\cdot)$. In addition to the fixed effects, $\boldsymbol{z}_{Sr}$ represents a vector of binary covariates associated to random effects $\boldsymbol{\gamma} \sim \mathcal{N}(0, \Sigma)$. These random effects account for unobserved heterogeneity, such as variation in the average hazard rates of interactions associated to different classrooms,  sender groups or receivers. 

The next section describes a number of endogenous covariates that may play a role in describing the dynamics of gossiping.

\subsection{Endogenous Gossiping Covariates}
The hyperevent nature of gossiping requires   covariates that account for its higher-order structure. Moreover, we focus on the case of multiple senders and a single receiver, rather than the case of a single sender and multiple receivers considered by \citet{LernerLomi2023} for modelling  email communications.
Table \ref{tab:covariates} lists a number of endogenous covariates that are of potential interest in describing the dynamics of gossiping. We split these into covariates that capture monadic effects, dyadic effects and those that refer to triadic effects. 

\begin{table}[t]
    \centering 
    \makebox[\linewidth][c]{%
  \resizebox{1\textwidth}{!}{ 

    \begin{tabular}{p{6cm} p{3.5cm} p{9.8cm}}       
        \toprule
        ~~\\
        \multicolumn{3}{c}{\textbf{\large endogenous covariates describing monadic effects}}\\  
        ~~\\
        \raisebox{-1em}{\textbf{Sender Degree}} & 
        \raisebox{-2em}{\begin{tikzpicture}[scale=0.4]
        \draw[thick, draw=black, fill=blue!50, opacity=0.3] (1.5, 1.5) circle (0.4);
        
        \draw[thick, draw=black, fill=blue!80, opacity=0.4] (4.5, 2.5) circle (0.3);
        \draw[thick, draw=black, fill=blue!80, opacity=0.4] (4.5, 1.5) circle (0.3);
        \draw[thick, draw=black, fill=blue!80, opacity=0.4] (4.5, 0.5) circle (0.3);
        
        \draw[thick, ->] (1.8, 1.8) -- (4.1, 2.5);
        \draw[thick, ->] (1.9, 1.5) -- (4.1, 1.5);
        \draw[thick, ->] (1.8, 1.2) -- (4.1, 0.5);
    
        \node at (1.5, 2.35) {$S$};
        \end{tikzpicture}}
        & \raisebox{-1em}{$x^{(\mathrm{sd})}_{S}(t)=\dfrac{1}{|S|} \displaystyle\sum_{\substack{t_i < t}} 1_{\{ S \subseteq S_i\}}$}\\
        \vspace{-0.8cm} \textbf{Receiver Degree} & 
        \begin{tikzpicture}[scale=0.4]
    \draw[thick, draw=black, fill=blue!80, opacity=0.4] (4.5, 1.5) circle (0.3);
    
    \draw[thick, draw=black, fill=blue!50, opacity=0.3] (1.5, 2.5) circle (0.4);
    \draw[thick, draw=black, fill=blue!50, opacity=0.3] (1.5, 1.5) circle (0.4);
    \draw[thick, draw=black, fill=blue!50, opacity=0.3] (1.5, 0.5) circle (0.4);
    
    \draw[thick, ->] (1.9, 2.5) -- (4.2, 1.8);
    \draw[thick, ->] (1.9, 1.5) -- (4.2, 1.5);
    \draw[thick, ->] (1.9, 0.5) -- (4.2, 1.2);

    \node at (4.5, 2.2) {$r$};
\end{tikzpicture}
         & \vspace{-0.8cm} $x^{(\mathrm{rd})}_{r}(t)=\displaystyle\sum_{\substack{t_i < t}}  1_{\{r = r_i\}}$ \\
         \hline
         ~~\\
         \multicolumn{3}{c}{\textbf{\large endogenous covariates describing dyadic effects}}\\ ~~\\
        \vspace{-0.5cm} \textbf{Repetition} & 
        \begin{tikzpicture}[scale=0.4]
        \draw[thick, draw=black, fill=blue!50, opacity=0.3] (1.5, 1.5) circle (0.7);
        
        \draw[thick, draw=black, fill=blue!80, opacity=0.4] (4.5, 1.5) circle (0.3);

        \draw[thick, ->] (2.2, 1.3) to[out=0, in=180] (4.2, 1.3);
        \draw[thick, ->, dashed] (2.2,1.6) to[out=0, in=180] (4.2, 1.6);

        \node at (1.5, 2.7) {$S$};
        \node at (4.5, 2.2) {$r$};
        \end{tikzpicture}
        & \vspace{-0.8cm} $x_{Sr}^{(\mathrm{rep})}(t) = \dfrac{1}{|S|} \displaystyle\sum_{\substack{t_i < t}} 1_{\{S_i = S \land r_i = r\}}$\\
        \textbf{Subset Repetition} &  
        \raisebox{-0.3cm}{\begin{tikzpicture}[scale=0.4]
        \draw[thick, draw=black, fill=blue!50, opacity=0.3] (1.5, 1.5) circle (1);
        
        \draw[thick, draw=black, fill=blue!80, opacity=0.4] (4.5, 1.5) circle (0.3); 

        \begin{scope}[fill opacity=0.3]
            \filldraw[fill=blue!60] 
            (1.1,1.7) 
            to[out=90,in=180] (1.6,2.2)  
            to[out=0,in=90] (2.1,1.5)
            to[out=270,in=0] (1.8,0.9)
            to[out=180,in=270] (1.1,1.7);
        \end{scope}

        \draw[thick, ->] (2.1, 1.3) to[out=0, in=180] (4.2, 1.3);
        \draw[thick, ->, dashed] (2.5,1.6) to[out=0, in=180] (4.2, 1.6);

        \node at (1.5, 3) {$S$};
        \node at (4.5, 2.2) {$r$};        
        \node at (1.6, 1.6) {\(\scriptstyle S_i\)};
        \end{tikzpicture}} & 
        \vspace{-0.8cm} \(x_{Sr}^{(\mathrm{sub\_rep})}(t)= \displaystyle\sum_{p=1}^{|S|} \frac{1}{\binom{|S|}{p}}\sum_{S' \in \binom{S}{p}} \mathrm{hy\_deg}_{t}(S', r)\)\\

        \vspace{-0.5cm} \textbf{Retaliation/Reciprocity} & 
        \begin{tikzpicture}[scale=0.4]
        \draw[thick, draw=black, fill=blue!50, opacity=0.3] (1.5, 1.5) circle (1);

        \draw[thick, draw=black, fill=blue!50, opacity=0.3] (4.5, 1.5) circle (1);
        
        \draw[thick, draw=black, fill=blue!80, opacity=0.4] (4.5, 1.6) circle (0.3); 

        \draw[thick, draw=black, fill=blue!80, opacity=0.4] (1.7, 1.3) circle (0.3);    

        \draw[thick, ->, dashed] (2.5,1.6) to[out=0, in=180] (4.1, 1.6);
        \draw[thick, ->]  (3.5, 1.3)  to[out=180, in=0] (2.1, 1.3);

        \node at (1.5, 3) {$S$};
        \node at (4.5, 3) {$S_i$};
        \node at (5, 1.9) {\(\scriptstyle r\)};
        \node at (1.3, 1.8) {\(\scriptstyle r_i\)};
        \end{tikzpicture}
        & \vspace{-0.8cm} $x^{(\mathrm{rec})}_{Sr}(t)=\dfrac{1}{|S|}\displaystyle\sum_{t_i < t} 1_{\{r \in S_i \, \wedge \, r_i \in S\}}$   \\
        \hline
        \\
        \multicolumn{3}{c}{\textbf{\large endogenous covariates describing triadic effects}}\\  
        ~~\\
        \vspace{-0.8cm}
        \textbf{Transitive Closure} &
        \begin{tikzpicture}[scale=0.4]
        \draw[thick, draw=black, fill=blue!50, opacity=0.3] (0, 0) circle (0.7);
        \draw[thick, draw=black, fill=blue!50, opacity=0.3] (1.5, 2.5) circle (0.7);
        \draw[thick, draw=black, fill=blue!80, opacity=0.4] (3, 0) circle (0.3);    
        \draw[thick, draw=black, fill=blue!80, opacity=0.4] (1.5, 2.3) circle (0.3);
    
        \node at (-0.9, 0.8) {$S$};
        \node at (3.6, 0.4) {$r$};
        \node at (1.5, 2.8) {\(\scriptstyle a\)};
    
        \draw[thick, ->] (2, 2) -- (2.8, 0.3); 
        \draw[thick, ->] (0.3, 0.6) -- (1.2, 2.1); 
    
        \draw[thick, dashed, ->] (0.7, 0) to[out=0, in=180] (2.6, 0); 
         \end{tikzpicture}
        & \vspace{-1cm} $x^{(\mathrm{tc})}_{Sr}(t)= \dfrac{1}{|S|} \displaystyle \sum_{\substack{a \neq s, r \\ s \in S}} \min \{\mathrm{hy\_deg}_t(S, a), \,  \mathrm{hy\_deg}_t(a, r)\} $   \\
        \vspace{-1cm}
        \textbf{Cyclic Closure} &
        \begin{tikzpicture}[scale=0.4]
        \draw[thick, draw=black, fill=blue!50, opacity=0.3] (0, 0) circle (0.7);
        \draw[thick, draw=black, fill=blue!50, opacity=0.3] (1.5, 2.5) circle (0.7);
        \draw[thick, draw=black, fill=blue!80, opacity=0.4] (3, 0) circle (0.3);    
        \draw[thick, draw=black, fill=blue!80, opacity=0.4] (1.5, 2.3) circle (0.3);
    
        \node at (-0.9, 0.8) {$S$};
        \node at (3.6, 0.4) {$r$};
        \node at (1.5, 2.8) {\(\scriptstyle a\)};
    
        \draw[thick, ->] (2.8, 0.3) -- (1.8, 2.1); 
        \draw[thick, ->] (1.1, 1.9) -- (0.3, 0.7); 
    
        \draw[thick, dashed, ->] (0.7, 0) to[out=0, in=180] (2.6, 0); 
         \end{tikzpicture}
         & \vspace{-1cm} $x^{(\mathrm{cc})}_{Sr}(t)= \dfrac{1}{|S|} \displaystyle \sum_{\substack{a \neq s, r \\ s \in S}} \min \{\mathrm{hy\_deg}_t(r, a) , \,\mathrm{hy\_deg}_t(a, s) \}$   \\
        \vspace{-1cm}
        \textbf{Sender Balance} &
        \begin{tikzpicture}[scale=0.4]
        \draw[thick, draw=black, fill=blue!50, opacity=0.3] (0, 0) circle (0.7);
        \draw[thick, draw=black, fill=blue!50, opacity=0.3] (1.5, 2.5) circle (0.7);
        \draw[thick, draw=black, fill=blue!80, opacity=0.4] (3, 0) circle (0.3);    
        \draw[thick, draw=black, fill=blue!80, opacity=0.4] (1.5, 2.3) circle (0.3);
    
        \node at (-0.9, 0.8) {$S$};
        \node at (3.6, 0.4) {$r$};
        \node at (1.5, 2.8) {\(\scriptstyle a\)};
    
        \draw[thick, ->] (2, 2) -- (2.8, 0.3); 
        \draw[thick, ->] (1.1, 1.9) -- (0.3, 0.6); 
    
        \draw[thick, dashed, ->] (0.7, 0) to[out=0, in=180] (2.6, 0); 
         \end{tikzpicture}
        & \vspace{-1cm} $x^{(\mathrm{sb})}_{Sr}(t) = \dfrac{1}{|S|}\displaystyle \sum_{\substack{a \neq s, r \\ s \in S}} \min \{ \mathrm{hy\_deg}_{t}(a, s), \, \mathrm{hy\_deg}_{t}(a, r) \}$   \\
        \vspace{-1cm}
        \textbf{Receiver Balance} & 
        \begin{tikzpicture}[scale=0.4]
        \draw[thick, draw=black, fill=blue!50, opacity=0.3] (0, 0) circle (0.7);
        \draw[thick, draw=black, fill=blue!50, opacity=0.3] (1.5, 2.5) circle (0.7);
        \draw[thick, draw=black, fill=blue!80, opacity=0.4] (3, 0) circle (0.3);    
        \draw[thick, draw=black, fill=blue!80, opacity=0.4] (1.5, 2.3) circle (0.3);
    
        \node at (-0.9, 0.8) {$S$};
        \node at (3.6, 0.4) {$r$};
        \node at (1.5, 2.8) {\(\scriptstyle a\)};
    
        \draw[thick, ->] (2.8, 0.2) -- (1.6, 1.98); 
        \draw[thick, ->] (0.3, 0.6) -- (1.4, 1.95); 
    
        \draw[thick, dashed, ->] (0.7, 0) to[out=0, in=180] (2.6, 0); 
         \end{tikzpicture}
         & \vspace{-1cm} $x^{(\mathrm{rb})}_{Sr}(t)  = \dfrac{1}{|S|}\displaystyle \sum_{\substack{a \neq s, r \\ s \in S}} \min \{ \mathrm{hy\_deg}_{t}(S, a), \, \mathrm{hy\_deg}_{t}(r, a) \}$   \\
        \bottomrule
    \end{tabular}}}
    \caption{Endogenous covariates describing gossiping hyperevents based on the history of the relational process. Solid lines ($\to$) refer to past relational hyperevents, while dashed arrows ($\dashrightarrow$) indicate current relational hyperevents.}
    \label{tab:covariates}
\end{table}

\paragraph{Sender Degree.} This covariate measures the activity level of a sender group. We define it as the number of times a sender set $S$ is involved in a gossiping hyperevent before the current time $t$. As this depends on the size of the sender set, we normalize it by the cardinality of $S$, leading to the definition
\begin{equation*}
    x^{(\mathrm{sd})}_{S}(t) = \dfrac{1}{|S|} \sum_{t_i \, < \, t} 1_{\{ S \, \subseteq \,  S_i \}}.
\end{equation*}
In gossip dynamics, a high sender degree indicates a persistent gossiping activity associated to a group of gossipers, so an effect associated to this variable may point to the tendency of individuals to maintain their role as central gossipers \citep{sparsenet}.

\paragraph{Receiver Degree.} Similarly, the receiver degree captures how often a given target has been the subject of gossiping prior to the current time $t$, i.e., 
\begin{equation*}
    x^{(\mathrm{rd})}_{r}(t) = \sum_{t_i \, < \, t} 1_{\{ r \, = \, r_i \}}.
\end{equation*}
For example, Figure~\ref{fig:hoi} shows the same receiver being the target of gossiping at time $t_2$ and at time $t_0$. In a school setting, this is an important variable for assessing whether bullying is taking place \citep{KisfalusiGossipReputation}. 

\paragraph{Repetition.} Similarly to \citep{LernerLomi2023}, but adapted to our setting with multiple senders, we include a covariate that measures past repetition of a hyperevent. 
Formally, this is defined as the number of past occurrences in which the same sender set $S$ has gossiped about a receiver $r$, normalized by the size of $S$:
\begin{equation*}
    x_{Sr}^{(\mathrm{rep})}(t) = \frac{1}{|S|} \sum_{t_i \, < \, t} 
        1_{\{ S_i \, = \, S \, \wedge \, r_i \, = \, r \}}.
\end{equation*}

\paragraph{Reciprocity (Retaliation).} 
A natural question in the study of gossiping is whether students are more likely to gossip about individuals who they believe are gossiping about them
\citep{nynkegossip}. 
This mechanism is naturally dynamic: retaliation occurs when a previous target of gossip becomes a sender and spreads information about those who gossiped about them. 
We operationalize this by counting whether the current receiver $r$ has previously acted as a sender gossiping about members of $S$:
\begin{equation*}
x^{(\mathrm{rec})}_{Sr}(t) \;=\; \frac{1}{|S|} \sum_{t_i \, < \, t} 1_{\{ r \, \in \, S_i \;\wedge\; r_i \,\in \, S \}}.
\end{equation*}
For example, Figure~\ref{fig:hoi}, shows a retaliation at time $t_1$ of the event that happened at time $t_0$ and at time $t_2$ of the event that happened at time $t_1$.

\paragraph{Subset Repetition.} 
This covariate relaxes the definition of repetition by measuring whether subsets of senders within $S$ have previously gossiped about $r$. In particular, for each possible subset of $S$ of size $p$,  we consider all subsets $S' \subseteq S$ of that size and, for each of them, we count how often they  have previously gossiped about $r$. Similarly to the definition of \citet{LernerLomi2023} for the case of multiple receivers, we define this last  quantity as the hyperdegree of $S'$ with respect to $r$, that is
\begin{equation}
\mathrm{hy\_deg}_{t}(S', r) = \sum_{t_i \, < \, t} 1_{\{ S' \, \subseteq \, S_i \, \land \, r \, = \, r_i \}}.
\label{eq:hyperdegree}
\end{equation}
These hyperdegrees are normalized by the number of subsets $S'$ of size $p$ and summed across all possible sizes $p = 1, \dots, |S|$, leading to the following definition of subset repetition
\begin{equation*}
    x_{Sr}^{(\mathrm{sub\_rep})}(t) = \sum_{p\, =\,1}^{|S|} \frac{1}{\binom{|S|}{p}} \sum_{S' \, \in \, \binom{S}{p}} \mathrm{hy\_deg}_{t}(S', r).
\end{equation*}
For example, Figure~\ref{fig:hoi}, shows a subset repetition at time $t_2$ of the event that happened at time $t_0$.

The notion of hyperdegrees serves as the basic building block for the definition of covariates that describe triadic effects, which have been extensively studied in the social networks and relational event literature \citep{rutaheterogeneity}. In particular, it allows us to extend notions of closure and balance from dyadic events to higher-order interactions. Below we describe a number of covariates that will be considered for modelling the potential triadic effects of gossiping.

\paragraph{Transitive Closure.} 
This covariate captures the intuition that friends of friends are likely to become gossip conduits \citep{giardini2019}. 
If a sender set $S$ has gossiped about some alter $a$, and $a$ together with someone else has gossiped about a receiver $r$, then  $S$ gossiping about $r$ completes the transitive path.  To capture this type of effect in our hypergraph framework, we define the following covariate 
\begin{equation*}
x^{(\mathrm{tc})}_{Sr}(t) \;=\; \frac{1}{|S|} 
\sum_{\substack{a \, \neq \, s, r \\ s \, \in \, S}} 
\min \Big\{ \text{hy\_deg}_t(S, a), \; \text{hy\_deg}_t(a, r) \Big\},
\end{equation*}
which keeps track of whether a sender set $S$ can reach a receiver $r$ through an intermediary $a$. Note that for the definition of this covariate, as well as all subsequent ones, we account for the direction of interactions, but not for their temporal order, i.e.,  we do not require that \( S \to a \) occurs prior to \( a \to r \).

\paragraph{Cyclic Closure.} 
Cyclic effects capture the circulation of gossiping along closed loops. 
If prior to time $t$, a receiver $r$ has gossiped about an intermediary node $a$, and $a$ has subsequently gossiped about a sender $s \in S$, then gossiping flowing back from $S$ to $r$ at time $t$ completes this cycle. Such patterns resonate with the gossip triangle identified in prior studies \citep{Ellwardt}.
\begin{equation*}
x^{(\mathrm{cc})}_{Sr}(t) \;=\; \frac{1}{|S|} 
\sum_{\substack{a \, \neq \, s, r \\ s \, \in \, S}} 
\min \Big\{ \text{hy\_deg}_t(r, a), \; \text{hy\_deg}_t(a, s) \Big\}.
\end{equation*}

\paragraph{Sender Balance.} 
This covariate quantifies the extent to which other actors direct gossip towards both a given sender and a given receiver. Thus it is defined by
\begin{equation*}
x^{(\mathrm{sb})}_{Sr}(t) \;=\; \frac{1}{|S|} 
\sum_{\substack{a \, \neq \, s, r \\ s \, \in \, S}} 
\min \Big\{ \text{hy\_deg}_{t}(a, s), \; \text{hy\_deg}_{t}(a, r) \Big\}.
\end{equation*}
Intuitively, this covariate captures situations in which a third party $a$ is simultaneously connected to $s$ and $r$. This resembles so called coalition triads \citep{wittek1998}, where gossip circulates more among actors with shared connections.

\paragraph{Receiver Balance.} 
This covariate measures the extent to which a sender set $S$ and a receiver $r$ target the same third parties with their gossiping. This is defined by
\begin{equation}
x^{(\mathrm{rb})}_{Sr}(t) \;=\; \frac{1}{|S|} 
\sum_{\substack{a \, \neq \, s, r \\ s \, \in \, S}} 
\min \Big\{ \text{hy\_deg}_{t}(S, a), \; \text{hy\_deg}_{t}(r, a) \Big\}.
\end{equation}
It captures shared focus on common alters, reflecting consensus-building processes within triads. This aligns with structural balance theory \citep{cartwright1956}, which predicts that imbalanced triads generate relational tension, and this is resolved by aligning attention or judgments towards common targets \citep{halevy2019}.

\section{Inference from partially observed relational event data}
\label{sec:inf_partiallydata}

As discussed earlier, survey data are collected at discrete time points, asking whether gossiping occurred during that time. No information is recorded about when it occurred or how many times. This means that only partial information is available on the relational events. In particular, the event times are interval-censored, since all we know is that they belong to the wave leading up to the survey, and the counts of the events are right-censored, since all we know is that at least one gossiping event occurred during that wave. In this section, we derive the  likelihood of these data, accounting for their censored nature.

\subsection{The likelihood under censoring}

We assume that there is an underlying relational hyperevent model~\eqref{hazard} which describes the dynamics of gossiping hyperevents. Let $K$ be the number of surveys conducted and  let \((t_{k-1},t_k]\), for $k=1,\ldots, K$, the corresponding time intervals. From the inhomogeneous Poisson counting process associated to the relational event process, the number of hyperevents involving a sender set $S$ gossiping about a receiver $r$ during wave \(k\) evolves according to
\begin{equation}
N_{Sr}(t_k) - N_{Sr}(t_{k-1})~|~\mathcal{H}_{t_{k-1}} \sim \text{Poisson}\left(\int_{t_{k-1}}^{t_k} \lambda_{Sr}(u) \, du\right),
\end{equation}
conditional on the history $\mathcal{H}$ of the process up to time \(t_{k-1}\). 
Denoting with $y_{Srk}^*$ the increments, $y_{Srk}^* = N_{Sr}(t_k) - N_{Sr}(t_{k{-}1})~|~\mathcal{H}_{t_{k-1}}$, what we observe are the right-censored counts, i.e.,
\begin{equation*}
y_{Srk} \, = \, 1_{\{ y_{Srk}^* > 0 \}} \, = \,
\begin{cases}
1 & \text{if at least one hyperevent occured in } (t_{k{-}1}, t_k], \\
0 & \text{otherwise}.
\end{cases}
\end{equation*}
In other words, the count data are right-censored at $1$. This leads to the following likelihood
\begin{equation}\label{eq:likelihood}
L(\boldsymbol{\beta}) = \prod_{k=1}^K \prod_{(S,r) \in \mathcal{R}(k)} 
\underbrace{\left( e^{-\int_{t_{k-1}}^{t_k} \lambda_{Sr}(u) \, du} \right)^{1 - y_{Srk}}}_{\raisebox{-0.7em}{\footnotesize $y^*_{Srk} \, = \, 0$}} 
\underbrace{\left(1 - e^{-\int_{t_{k-1}}^{t_k} \lambda_{Sr}(u) \, du}\right)^{y_{Srk}}}_{\raisebox{-0.7em}{\footnotesize $y^*_{Srk} \, > \, 0$}},
\end{equation}
with $\mathcal{R}(k)$ denoting the risk set of potential hyperevents that could occur in wave $k$. Of those, the hyperevents that do no occur  contribute to the likelihood with a Poisson probability of zero, whereas the hyperevents that do occur contribute with a Poisson probability of a positive count, which is all the information we have from the surveys.

As the integrals in \eqref{eq:likelihood} are intractable, we approximate these by
\begin{equation*} \int_{t_{k-1}}^{t_k} \lambda_{Sr}(u) \, du \, \,  = \, \, (t_k - t_{k-1}) \, \lambda_{Sr}(\bar{t}), 
\end{equation*}
with $\bar{t}_k=(t_{k-1}+t_{k})/2$ the midpoint of the $(t_{k-1},t_k]$ interval. The relational hyperevent model~\eqref{hazard} provides the link between the intensities $\lambda_{Sr}(\bar{t})$ and the covariates $\boldsymbol{x}_{Sr}(\bar{t})$. 
Since the covariates and the hyperevents are recorded only at the point of the survey and not within the interval, we have various options on how to evaluate $\lambda_{Sr}(\bar{t})$. In the simulation study, we show how a good strategy is to consider 
\begin{equation}
\lambda_{Sr}(\bar{t}_k) \, = \, Y_{Sr}(\bar{t}_k) \, \lambda_0(\bar{t}_k) \,
\exp\!\left\{
f\Big(\dfrac{\boldsymbol{x}_{Sr}(t_{k-1})+\boldsymbol{x}_{Sr}(t_{k})}{2}\Big) + \boldsymbol{\gamma}^\top \boldsymbol{z}_{Sr}
\right\},
\label{eq:lambdatstar}
\end{equation}
i.e., to evaluate the hazard using the average of the covariate values at the two extremes of the interval. Indeed, this value is expected to carry most of the information about the time-varying behaviour of the variables within the interval.

Substituting the intensities from equation~\eqref{eq:lambdatstar} into the likelihood from equation~\eqref{eq:likelihood}, leads to the approximation
\begin{equation}
\bar{L}(\boldsymbol{\beta}) = 
\prod_{k=1}^K \prod_{(S,r) \in \mathcal{R}(k)} 
\left[
e^{- (t_k - t_{k-1}) \lambda_{Sr}(\bar{t}_k)}
\right]^{1 - y_{Srk}}
\left[
1 - e^{- (t_k - t_{k-1}) \lambda_{Sr}(\bar{t}_k)}
\right]^{y_{Srk}},
\label{eq:full_like}
\end{equation}
which is the likelihood of a right-censored Poisson regression model with the quantities $\log(t_k - t_{k-1})$ as offsets. Indeed, conditional on the covariates, the rate $\mu_{Srk}$ of hyperevent $(S,r)$ in the risk set of wave $k$ is described by
\[\log(\mu_{Srk}) = \log(t_k - t_{k-1}) + \log \lambda_{Sr}(\bar{t}).\]

Since the censoring is at $1$, this censored Poisson regression is equivalent to a Binomial regression with complementary log-log (cloglog) link. Indeed,
\begin{equation*}
\pi_{Srk}=P( y_{Srk} = 1 ) = P(y^*_{Srk}>0)= 1- P(y^*_{Srk} = 0) = 1 - e^{-\mu_{Srk}},
\end{equation*}
leading to 
\[ \log(-\log(1-\pi_{Srk}))=\log(\mu_{Srk}),\]
and thus to the predictor
\[ \log(-\log(1-\pi_{Srk}))=\log(t_k - t_{k-1}) + \log \lambda_{Sr}(\bar{t}).\]
So the binary vector of gossip indicators $y_{Srk}$ can be equivalently modelled via a Binomial regression model with offsets $\log(t_k - t_{k-1})$ and using the complementary log-log link function.

\subsection{Efficient inference of flexible RHEMs}

A key advantage of the likelihood formulation derived in the previous section is that efficient implementations of flexible models are available in standard statistical software packages, particularly for Binomial complementary log-log (cloglog) regression models. In particular, the complex dynamics of gossiping necessitates the use of flexible generalized additive modelling with fixed, smooth and random effects. These effects are implemented in the \texttt{R} package \texttt{mgcv} \citep{wood2017} both for the case of censored Poisson as well as for Binomial cloglog regression. The latter will be used for the analyses in the next sections.

\section{Simulation Study}
\label{sec:simulation}

In this section, we evaluate the effectiveness of the proposed approach through a simulation study. In a first study, we evaluate the quality of parameter estimation using the proposed approach in the presence of partially observed data. In a second simulation study, we compare different ways of evaluating the time-varying covariates that are included in the model. These, as all the other available data, are measured only at the beginning and at the end of a wave.

\subsection{Parameter estimation from partially observed data}\label{sec:simu1}

We evaluate the performance of our proposed inferential procedure by constructing partially observed right-censored relational data, as in the gossiping application motivating the methodological development. In particular, we generate data from a relational hyperevent model characterized by the following intensity process
\begin{equation}
    \lambda_{Sr}(t)=\lambda_0 \exp(  
    \beta_1 x_{r}^{(\mathrm{girl\_alter})}+
    f(x_{Sr}^{\left(\text{age}\right)})),
\label{eq:model1}
\end{equation}
where we mimic the effect of two exogenous covariates, gender and age, on the dynamics of gossiping. To this end, we define $x_{r}^{(\mathrm{girl\_alter})}$ as a binary indicator taking the value of $1$ if the receiver is female, and $0$ otherwise, i.e.,
\begin{equation*}
x_{r}^{(\mathrm{girl\_alter})} \;=\; 1_{\{\text{gender}(r) \, = \,\text{female}\}},
\end{equation*}
and set $\beta_1$ to $0.9$ to indicate the tendency for gossiping to be directed more toward women than men. 
The second variable, $x_{Sr}^{(\mathrm{age})}$,  is instead defined as the average age of all senders and receivers in a hyperevent, i.e., 
\begin{equation*}
x_{Sr}^{(\mathrm{age})} \;=\; \frac{1}{|S| + 1} \Bigg( \sum_{s \, \in \, S} \text{age}(s) + \text{age}(r) \Bigg).
\end{equation*}
For this variable, we simulate a decreasing smooth effect  on the hazard of gossiping by defining
\begin{equation*}
    f(x_{Sr}^{(\mathrm{age})})= \frac{1}{1+\exp(2 (x_{Sr}^{(\mathrm{age})}-16))}.
\end{equation*}

We generate relational hyperevents among $8$ interacting students from this process  via a Gillespie algorithm \citep{Gill77}. In particular, inter-arrival times are drawn from an exponential distribution, and at each event time the occurring hyperevent is sampled from the risk set according to a multinomial distribution. We  assume that, at any time point,  all events between a sender group of size at most $3$ and a receiver among the remaining students are at risk of happening. Given the complete data, we consider $6$ waves defined by the intervals ($k-1,k]$, $k=1,\ldots,6$, and construct a right-censored version of these data, by considering for each hyperevent whether it happened at least once within each wave.

We use the proposed inferential procedure to recover the intensity of the process from the censored data. In particular, we fit a Binomial cloglog model using the \texttt{gam} function from the R package \texttt{mgcv}, with a thin-plate regression spline on age \citep{wood2003}. 
The simulations are performed under  a small value of $\lambda_0$ ($\lambda_0 = 0.25$) and a large value ($\lambda_0 = 0.75$), creating in this way two settings with a varying average number of occurrences for each hyperevent. In particular, a larger $\lambda_0$ value will generate a larger number of occurrences of each hyperevent within each wave and therefore a larger  information loss when complete data are replaced by censored data, compared to a smaller value of $\lambda_0$, where censored data may be not too different to the complete data. For each setting, we perform 20 simulations of the process and summarize the results in terms of quality of parameter estimation of the linear (gender) and smooth (age) effects. 

\begin{figure}[tp]
        \centering
        \includegraphics[width=\linewidth]{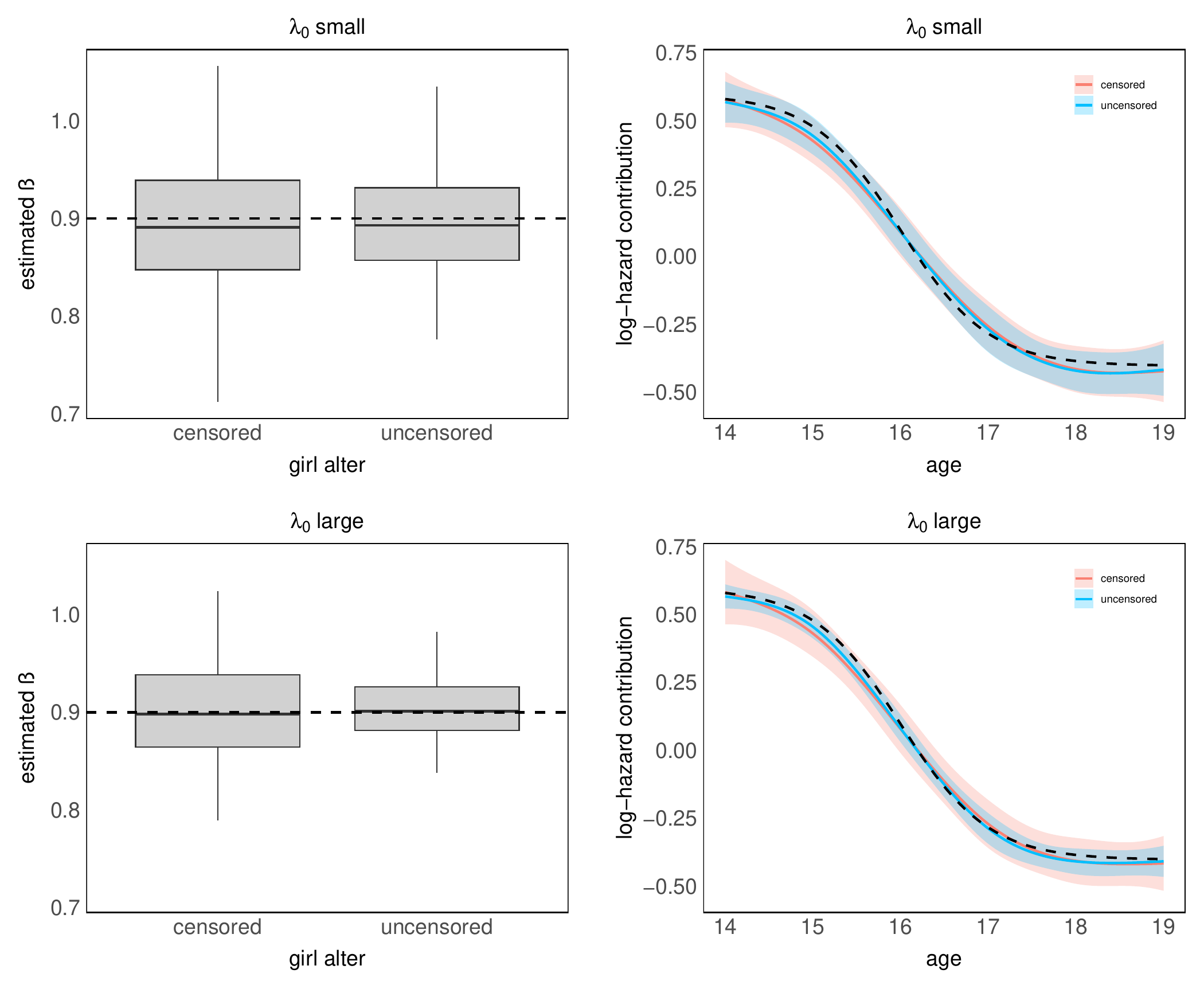}
        \begin{minipage}[t]{0.55\linewidth}
        \centering
        (a)
    \end{minipage}%
    \hfill
    \begin{minipage}[t]{0.41\linewidth}
        \centering
        (b)
    \end{minipage}
    \caption{
Simulating $100$ datasets of right-censored gossiping hyperevents among $8$ students across $6$ waves according to model~\eqref{eq:model1} with a small (top) and large (bottom) $\lambda_0$ value. True effects are represented by a dashed line. (a) Boxplots of the estimated $\beta_1$ effect of $x_{r}^{(\mathrm{girl\_alter})}$ on the hazard  and (b) 95\% confidence band of the smooth effects of $x_{Sr}^{(\mathrm{age})}$, centered at zero,  are well-calibrated under all settings when using  a Binomial cloglog \texttt{gam} model on censored data, while they have a higher uncertainty compared to the estimates from the complete uncensored data when the information loss is large (bottom). 
}\label{fig:linear_smooth}
\end{figure}

Figure~\ref{fig:linear_smooth} reports the results. 
In particular, Figure~\ref{fig:linear_smooth}a
shows boxplots of the estimated $\beta_1$ coefficient associated to the covariate $x_{r}^{(\mathrm{girl\_alter})}$ , under a small (top) and large (bottom) $\lambda_0$ value. Similarly, Figure~\ref{fig:linear_smooth}b reports the estimated smooth effects for the covariate $x_{Sr}^{(\mathrm{age})}$ under both settings. As a benchmark, we take the estimates obtained from the complete data via a Poisson generalized additive model with the same specifications as the Binomial cloglog model. The results show how the estimates from the censored data are well calibrated under all settings. As expected, the information loss due to censoring leads to an increase in the uncertainty of the estimates, particularly for the case of a large $\lambda_0$ where multiple occurrences of the same hyperevent within each wave are more likely. This is reflected by a wider boxplot in   Figure~\ref{fig:linear_smooth}a (bottom) and a wider confidence band in Figure~\ref{fig:linear_smooth}b (bottom), compared to the estimates from the uncensored data.

\subsection{Evaluating time-varying covariates from partially observed data}\label{sec:simu2}

Besides censoring on the number of gossiping  hyperevents, a second, connected, source of information loss is given by the fact that the event times are interval-censored, as their exact times are bounded within the interval of the corresponding wave. This form of information loss occurs frequently in relational event settings, where data are sometimes provided in the form of aggregate counts  across discrete time intervals rather than instantaneous events. 

In these cases, also the covariates that are used to describe the process are typically evaluated only at the extremes of the interval. Referring back to the gossiping data, this is clearly the case for endogenous covariates, as they can change only when new events occur and this information is available only at the time when the survey is handed out. But this is the case also for endogenous covariates that are also measured during the survey. Since most of the covariates  are time-varying, a question of interest is how best to evaluate these covariates. In particular, given a wave $(t_{k-1},t_k]$, one could evaluate the covariates
\begin{enumerate}
    \item $\:$  At time $t_{k-1}$, i.e., using only information on the history of the process prior to the wave. We denote this strategy with \texttt{past};
    \item $\:$ At time $t_k$, i.e., using all information up to and including the current wave. We denote this strategy with \texttt{current};
    \item $\:$ As an arithmetic mean of the past and current values. We denote this strategy with \texttt{average}.
\end{enumerate}
Intuitively, the last strategy should be the most effective one, as it provides the best description of the entire behaviour of the covariate across the interval.

We design a simulation study to compare the three evaluation strategies. In particular, we generate relational hyperevents from a model characterized by the following intensity process 
\begin{equation}
\lambda_{Sr}(t) = \lambda_0  \exp(\beta x(t)),
\label{eq:model2}
\end{equation}
where the time-varying covariate is defined by $x(t) = \log(t+1)$,
the baseline intensity $\lambda_0$ is set to 0.038 and the regression coefficient $\beta$ to 0.8. 
As the covariate is time-varying, we  use the tau-leap algorithm \citep{tau-leap} to simulate relational hyperevents from this model. As before, we consider $8$ interacting students, sender sets with at most $3$ individuals, and generate data up to time $t=6$. We consider time intervals $(k-1,k]$, $k=1,\ldots,6$, construct right-censored hyperevent data as before and assume that the covariate is measured only at the extremes of each interval.

\begin{figure}[t]
    \centering
    \includegraphics[width=0.6\linewidth]{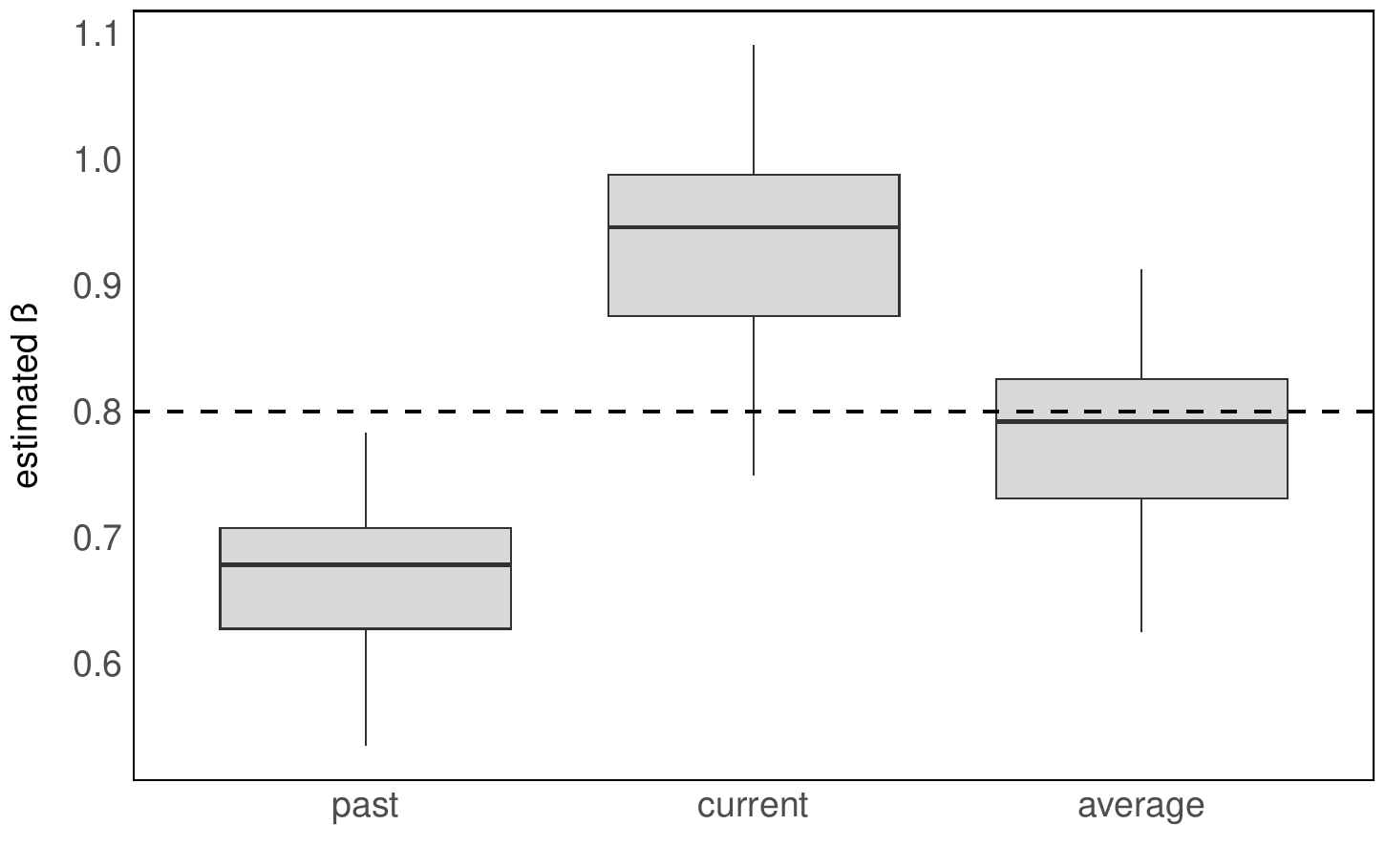}
    \caption{ Simulating $100$ datasets of right-censored gossiping hyperevents among $8$ students across $6$ waves according to model~\eqref{eq:model2} with a time-varying covariate. True effect $\beta$ is represented by a dashed line  The best estimation of the effect is obtained by a Binomial cloglog \texttt{gam} which uses the \texttt{average} evaluation of the covariate for each wave, rather than an evaluation at the beginning (\texttt{past}) or end (\texttt{current}) of each interval.
    }
    \label{fig:cov_evaluation}
\end{figure}

Figure \ref{fig:cov_evaluation} reports the results from $100$ simulations.
We consider the three different ways of evaluating the covariate at each time interval and, in each case, fit a Binomial cloglog \texttt{gam}, with a thin-plate regression spline on this covariate. The boxplots of the estimates of $\beta$ across the $100$ simulations and the three settings, show how, as expected, the \texttt{average} approach for the evaluation of the covariate leads to the best estimation of the parameters.

\section{Modelling gossiping in the RECENS school survey study}
\label{sec:dataanalysis}

\subsection{Gossiping hyperevent model}
We now return to the gossiping study, and consider the following relational hyperevent model to describe the dynamics of gossiping:
\begin{align*}
    \lambda_{Sr}(t)&=\lambda_0(t) 
      \exp \Big\{\beta_1 x_{r}^{(\mathrm{girl\_alter})} +
    f_1(x_{S}^{(\mathrm{girl\_ego})})+
    f_2(x_{S}^{(\mathrm{sd})}(t))+
    f_3(x_{r}^{(\mathrm{rd})}(t)) +
    f_4 (x_{Sr}^{(\mathrm{rep})}(t)) \\ & +
    f_5 (x_{Sr}^{(\mathrm{sub\_rep})}(t)) +
    f_6 (x_{Sr}^{(\mathrm{rec})}(t)) +
    f_7 (x_{Sr}^{(\mathrm{tc})}(t)) +
    f_8 (x_{Sr}^{(\mathrm{cc})}(t)) +
    f_{9} (x_{Sr}^{(\mathrm{sb})}(t)) +
    f_{10} (x_{Sr}^{(\mathrm{rb})}(t)) \\ & +
    \gamma_{\mathrm{class}(S,r)} +
    \gamma_{r} +
    \gamma_{S} \Big\}.
\end{align*}
Besides a flexible baseline function $\lambda_0$, capturing temporal changes on the rate of gossiping, the first two variables are exogenous variables defined by 
\begin{align*}
x_{r}^{(\mathrm{girl\_alter})} &= 1_{\{\text{gender}(r) \, = \,\text{female}\}},\\
x_{S}^{(\mathrm{girl\_ego})} &= \dfrac{1}{|S|}\sum_{s \in S} 1_{\{\text{gender}(s) \, = \,\text{female}\}},
\end{align*}
and describing the effect of gender on the rate of gossiping. The following $9$ variables are instead the endogenous covariates defined in Table~\ref{tab:covariates}, capturing the potentially viral nature of gossiping. The last three terms are random effects for class, receiver, and sender, respectively. These account for unobserved heterogeneity in the rate of gossiping between the different classes, the different receivers or the different sender groups, respectively. 
The inclusion of these random effects reduces potential confounding induced by unobserved heterogeneity. For example, \citet{rutaheterogeneity} show how failing to adjust for node degree heterogeneity may generate spurious ``ghost'' triadic effects. 
Besides the binary variable measuring the gender of the receiver $x_{r}^{(\mathrm{girl\_alter})}$, all other covariates are included in the model as flexible smooth effects, described by the functions \(\lambda_0, f_1, \dots, f_{10}\).

\subsection{The dynamics of gossiping inferred from the RECENS school survey study}
We fit the model on the school survey data discussed earlier. The four waves have different lengths, so we set the time intervals according to the duration in months of each wave, i.e., $(0, 2]$, $(2, 8]$, $(8,20)$, $(20,32]$, respectively.  The covariates are evaluated for all hyperevents in the risk set across all waves, so for a total of $299,419$ potential hyperevents. Sender and receiver degree are included with a log-transformation, $\log(x+1)$, for computational stability purposes. For the time-varying covariates, we take the average of their evaluation at the two extremes of the time interval of the corresponding wave, as indicated by the simulation study. We then fit a Binomial generalized additive mixed model with cloglog link  via  the \texttt{bam} function from the \texttt{mgcv} package, with a thin-plate regression spline on each smooth term  \citep{wood2017}. The observations from each wave have a corresponding offset given by the logarithm of the length in months of that wave, i.e., $2$, $6$, $12$, $12$ for the 4 waves, respectively. In order to perform automatic model selection, we use the double-penalty approach of \citet{penaltySelectTrue}, whereby each smooth term receives an additional penalty which allows it to be shrunk entirely to zero if it is uninformative. 

\begin{figure}[t]
    \centering
    \begin{minipage}[b]{0.4\linewidth}
        \centering        \includegraphics[width=\linewidth]{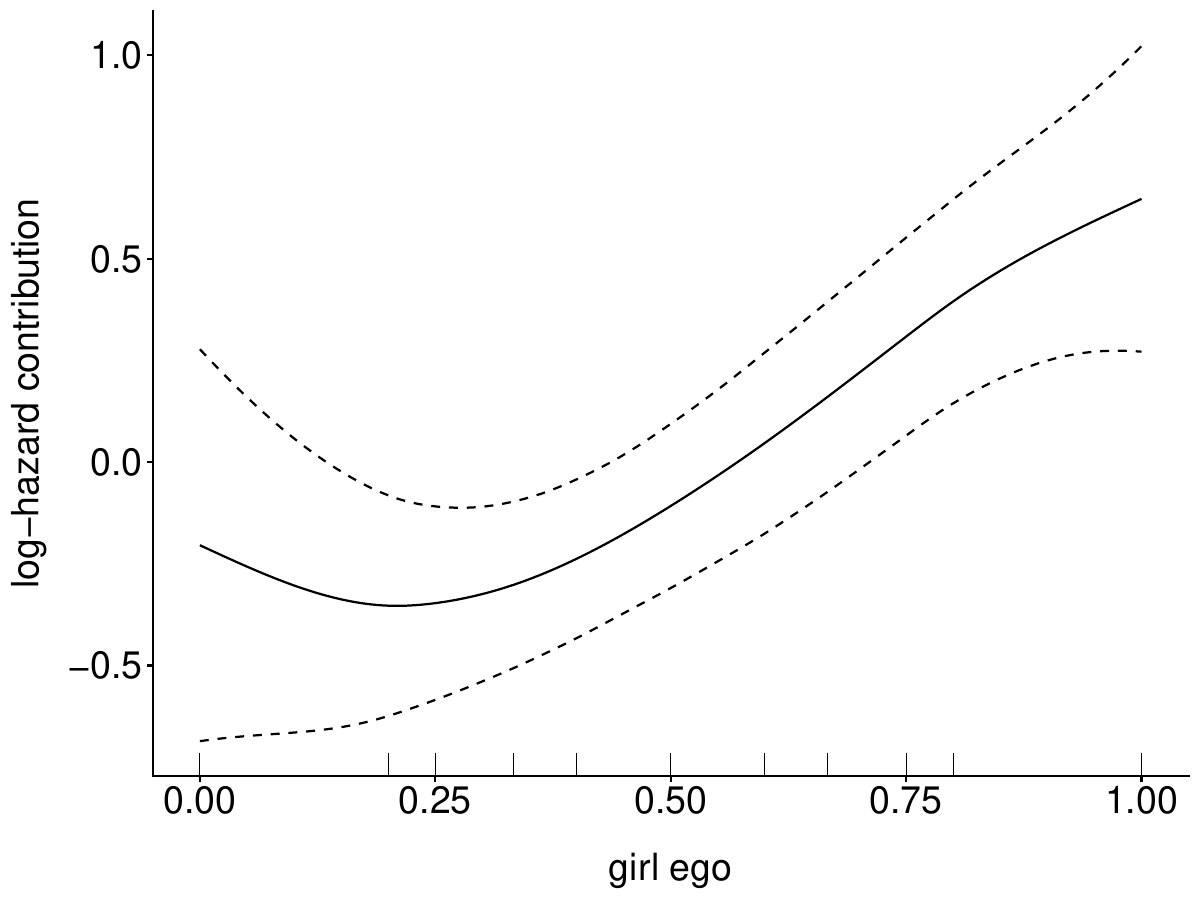}\\
(a)    
\end{minipage}
    \hfill
    \begin{minipage}[b]{0.4\linewidth}
        \centering
        \includegraphics[width=\linewidth]{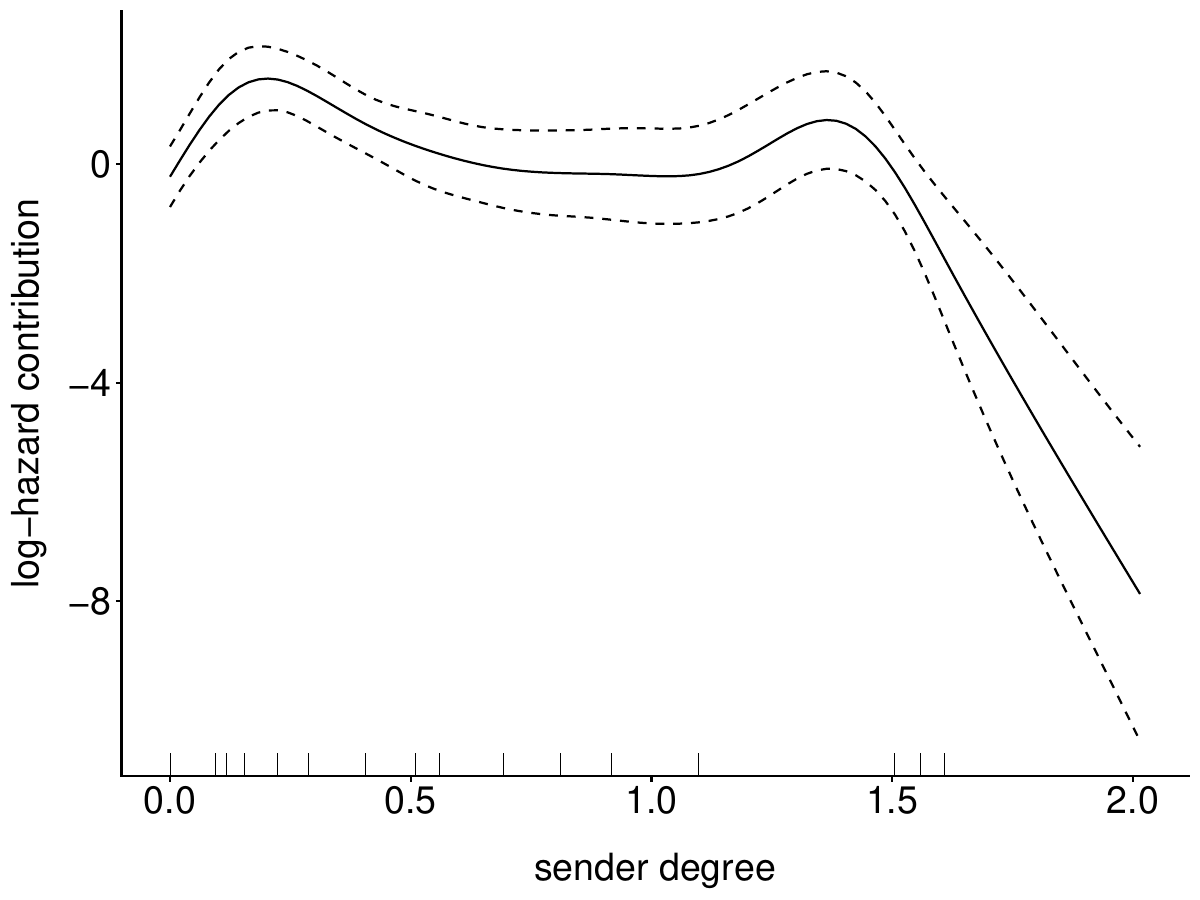}\\
(b)
    \end{minipage}  \hfill

    \makebox[\textwidth][c]{%
    \begin{minipage}[b]{0.4\linewidth}
        \centering
        \includegraphics[width=\linewidth]{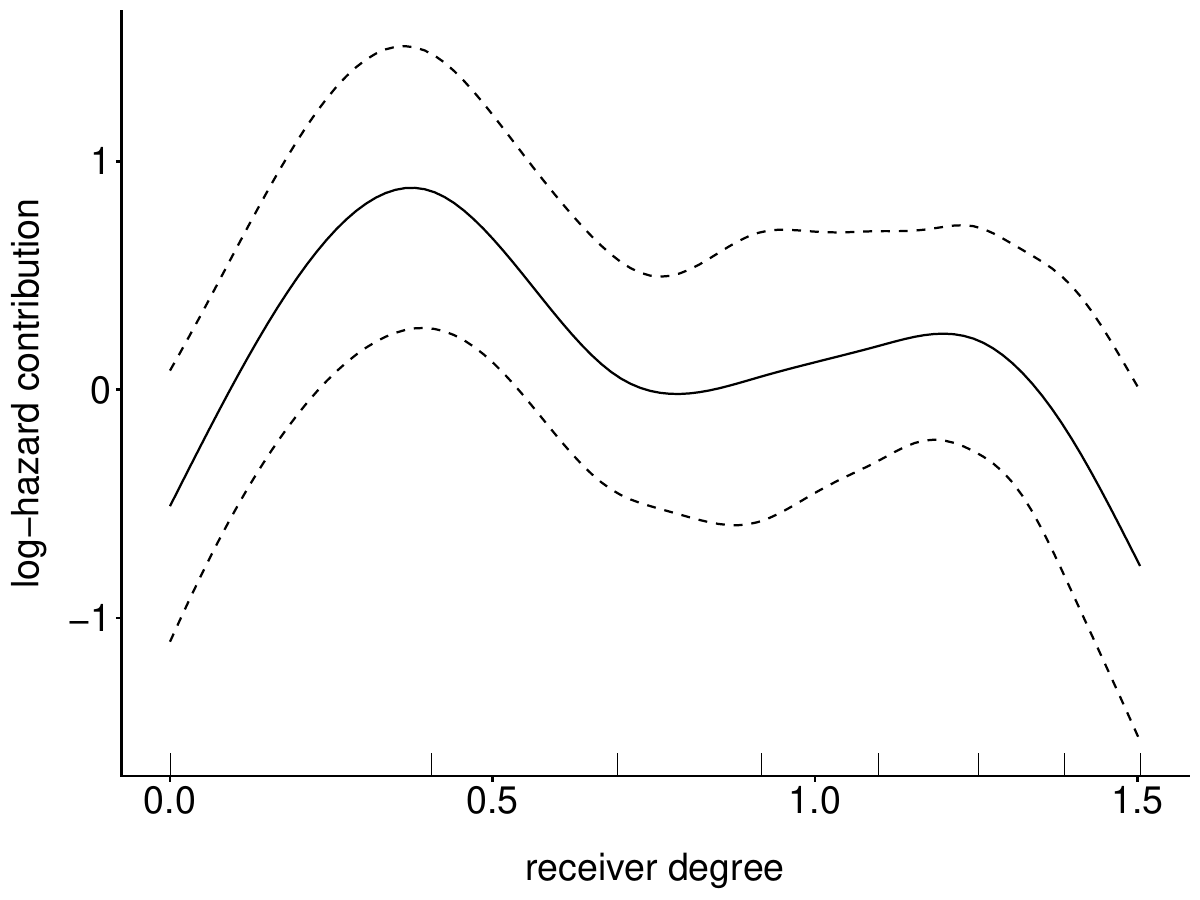} \\
(c)
    \end{minipage}
    \hfill
    \begin{minipage}[b]{0.4\linewidth}
        \centering            
        \includegraphics[width=\linewidth]{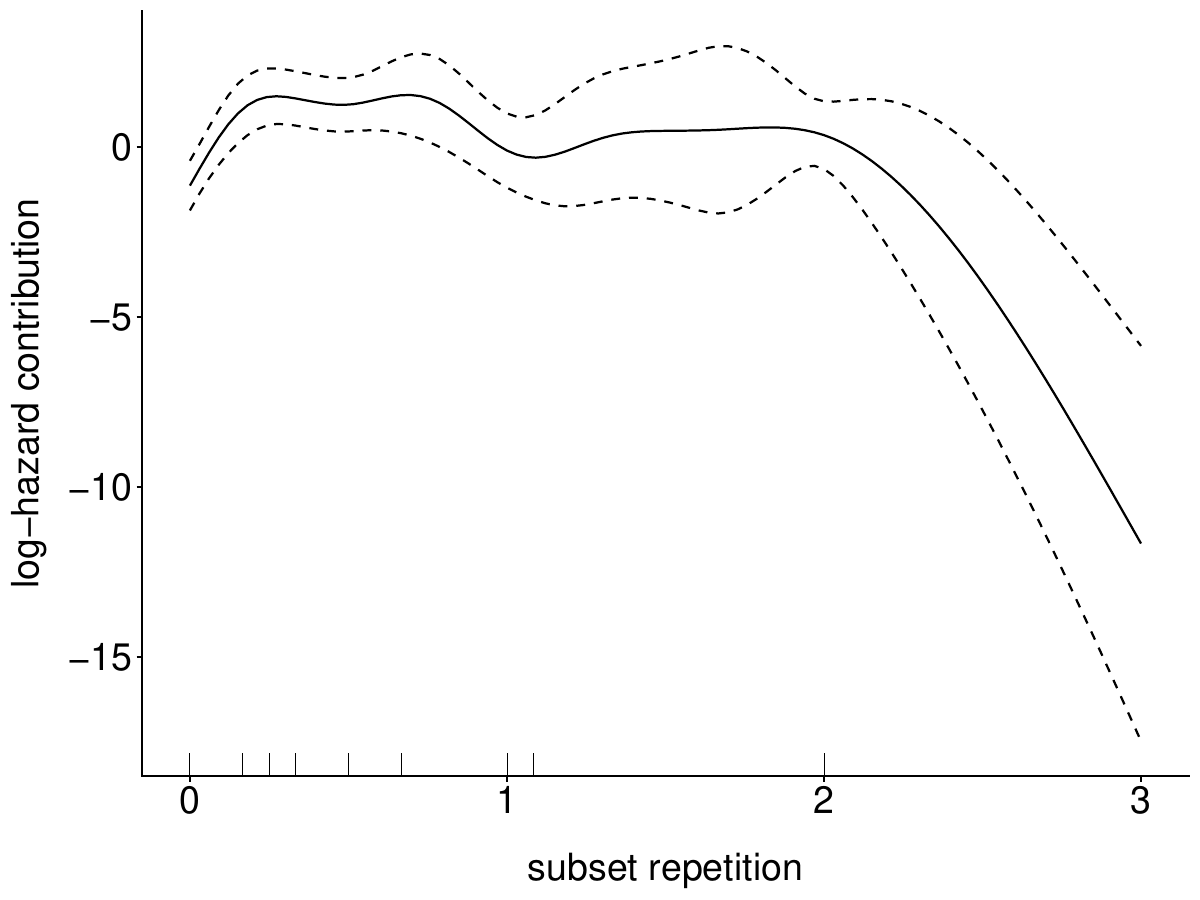} \\
(d)    \end{minipage}
    }

    \caption{Estimated smooth effects from the Binomial generalized additive mixed model of gossiping fitted to the school survey data shows how the rate of gossiping increases with (a) the proportion of girls in the sender set,  (b) the sender activity, (c) receiver activity and (d) the repetition of gossiping from subsets of the sender set. For (b)-(c)-(d) the increase is only at the beginning, indicating low levels of virality of gossiping. The solid line represents the estimated effect, while the dashed lines indicate the 95\% confidence interval.}
    \label{fig:smooth-effects}
\end{figure}

The results are summarized in Table~\ref{tab:gam_model}. Besides the fixed effects, which are found to be significant according to the small associated p-values, the informative variables are those with a large effective degrees of freedom (edf). Figure~\ref{fig:smooth-effects} plots the fitted smooth effects associated to four of the informative variables. The results highlight a number of key factors and mechanisms underlying gossiping.
\setlength{\tabcolsep}{15pt}
\renewcommand{\arraystretch}{1.5}
\begin{table}[t]
\resizebox{\textwidth}{!}{%
\begin{tabular}{lccccc}
\hline
\multicolumn{6}{l}{\textbf{Fixed effects}} \\
\hline
 & estimate & std. error & z value & p-value &  \\
\hline
intercept & -8.7922 & 0.4981 & -17.653 & $<2\times10^{-16}$ & *** \\
girl alter  & 0.5006  & 0.1858 &  2.694 & 0.00706 & ** \\
\hline
\multicolumn{6}{l}{\textbf{Smooth effects}} \\
\hline
 & edf & ref.df & chi-sq & p-value &  \\
\hline
girl ego                 & 1.634  & 9    & 70.06   & 0.000354 & *** \\
time                     & 0.971  & 3    & 98.82   & $<2\times10^{-16}$ & *** \\
sender degree   & 5.927  & 9    & 233.52  & $<2\times10^{-16}$ & *** \\
receiver degree   & 3.430  & 7    & 41.99   & 7.33e-05 & *** \\
repetition               & 5.737e-05 & 3 & 0.00    & 0.69478 &  \\
subset repetition        & 4.932  & 9    & 147.14  & $<2\times10^{-16}$ & *** \\
reciprocity              & 4.959e-05 & 9 & 0.00    & 0.80313 &  \\
transitive closure       & 7.853e-05 & 5 & 0.00    & 0.34624 &  \\
cyclic closure           & 3.822e-05 & 9 & 0.00    & 0.91670 &  \\
sender balance           & 6.748e-05 & 9 & 0.00    & 0.56604 &  \\
receiver balance         & 5.949e-05 & 7 & 0.00    & 0.49844 &  \\
\hline
\multicolumn{6}{l}{\textbf{Random effects}} \\
\hline
 & \# groups & std. dev. & 95\% CI  & \\
\hline
class     & 44 & 0.742 & [0.487, 1.130] &  \\
receiver  & 1686 & 0.497 & [0.186, 1.330] &  \\
sender set    & 7276 & 1.479 & [1.317, 1.662] &  \\
\hline
\end{tabular}%
}
\caption{Results from the Binomial generalized additive mixed model of gossiping fitted to the school survey data with a double-penalty approach \citep{penaltySelectTrue}, showing fixed, smooth and random effects. 
}
\label{tab:gam_model}
\end{table}

Firstly, girls are both more likely to gossip and to be the target of gossiping. The first aspect is supported by the increasing smooth effect of $x_{S}^{(\mathrm{girl\_ego})}$ (Figure~\ref{fig:smooth-effects}a), while the second aspect is reflected by the positive linear effect of $x_{r}^{(\mathrm{girl\_alter})}$ (Table~\ref{tab:gam_model}) on the rate of gossiping. Combined with the fact that the majority of sender groups is made up of only girls (52.3\%) this aligns with prior research suggesting that same-gender gossip is frequent, and is more common among girls \citep{nynkegossip,KisfalusiGossipReputation}.

Regarding the effect of the endogenous covariates, only a small number of these are found to play a significant role in gossiping. This may be partly due to the small number of time points available, which may limit the data available for estimating these effects,  in particular the triadic ones. As for the informative variables, the smooth effects relative to sender degree (Figure~\ref{fig:smooth-effects}b), receiver degree (Figure~\ref{fig:smooth-effects}c) and subset repetition (Figure~\ref{fig:smooth-effects}d) show a consistent non-linear pattern on the rate of gossiping, characterized by an initial increase, a plateau and a sharp decrease at high activity levels. This points to an intense but short-lived gossiping activity which does not become persistently viral. In particular, the fact that a sender group has been involved in a gossiping event, a receiver has been the target of a gossiping, or a sender group has targeted a specific receiver, increases the rate of the same sender group, the same target or the same event, respectively, to occur again in the future. However, this effect fades away after a few occurrences of the same event, and the event is in fact less likely to occur again at high activity levels. Whereas in the literature, it is suggested that students who gossip continue to do so \citep{KisfalusiGossipReputation}, our results show that this effect does not become viral.  Similarly, our results show how being the target of  past gossiping increases the rate of being targeted again, in line with the positive linear effect fitted by \cite{KisfalusiGossipReputation}. However, also in this case, the use of more complex models with non-linear smooth effects shows how there is no persistent viral targeting of the same receiver.

Another important aspect of our model, which is not considered by the existing literature on  gossiping, is the inclusion of random effects for sender group, receiver and class. Table~\ref{tab:gam_model} shows heterogeneity associated to each of these effects, as indicated by significant standard deviations. The use of these random effects allows us to disentangle the viral effect of gossiping from unobserved individual traits. The latter may be related to someone's personality, visibility, or social prominence, and may explain the high gossiping activity associated to some individuals or the persistent targeting of others. 
Our results indeed  suggest that being a gossiper or being targeted by gossiping is driven more by intrinsic traits than by past exposure, possibly explaining the apparent discrepancy with existing studies regarding the effect of endogenous covariates \citep{KisfalusiGossipReputation}.
Finally, our model is fitted to the data from all classes, rather than separately for a small selection of classes \citep{KisfalusiGossipReputation}. The joint modelling approach allows for borrowing strength across classes, which is particularly important in this study with a small number of total observations. However, the inclusion of a class random effect, allows us to conclude that there is significant heterogeneity at the class level which is accounted for by our model.

\section{Conclusion}

Gossiping is a social behaviour influenced by multiple factors in a complex and dynamic way. In this paper, we present a relational hyperevent model that describes the complexity of this process via a number of exogenous and endogenous variables. These variables capture  the heterogeneity and potential virality of the phenomenon and have been defined to account for the higher-order nature of gossiping, whereby at least two gossipers are needed to gossip about a third non-present person.

We fit the model to data from a longitudinal school survey from Hungarian secondary schools. Limitations in data collection require an extension of the inference to accommodate for right-censored hyperevent data and interval-censored event times. Thanks to a reformulation of the likelihood to that of standard regression models, we are  able to fit complex models with linear, smooth and random effects in the presence of partially observed data.  Our results show how the inclusion of smooth effects, capturing complex non-linear dynamics, and random effects, capturing unobserved heterogeneity, may play an important role in disentangling the viral effects of gossiping from those related to individual traits.

Overall, by adapting relational hyperevent models to higher-order gossiping interactions and extending them to handle partially observed data, our study broadens the applicability of these models to a wider range of empirical settings, providing researchers with a flexible tool for investigating collective dynamics from incomplete network data.

\section{Code availability}

The code for reproducing the simulation study and the empirical analysis of gossip data can be
found at the \texttt{GitHub} repository page https://github.com/veronicapoda/gossip-rhem.git.

\section*{Acknowledgments}

We thank the Research Center for Educational and Network Studies (RECENS) for providing access to the school survey dataset and useful information about the data.

\bibliographystyle{chicago}  
\bibliography{bib.bib}  

\end{document}